\documentclass[a4paper,fleqn,usenatbib]{mnras}

\usepackage{newtxtext,newtxmath}
\usepackage{subfig}
\usepackage[T1]{fontenc}
\usepackage{ae,aecompl}

\usepackage{graphicx}	% Including figure files
\usepackage{amsmath}	% Advanced maths commands
\usepackage{amssymb}	% Extra maths symbols

\title[Radio jets in NGC\,4151]{Radio jets in NGC\,4151: where eMERLIN meets HST}

\author[D.R.A. Williams et al.]{D.R.A. Williams,$^{1}$\thanks{E-mail: D.R.A.Williams@soton.ac.uk}
I.M. McHardy,$^{1}$
R.D. Baldi,$^{1}$
R.J. Beswick,$^{2}$
M.K. Argo,$^{2,3}$
\newauthor{B.T. Dullo,$^{4,5,6}$
J.H. Knapen,$^{5,6}$
E. Brinks,$^{7}$
D.M. Fenech,$^{8}$
C. Mundell,$^{9}$
}
\newauthor{T.W.B. Muxlow,$^{2}$
F. Panessa,$^{10}$
H. Rampadarath,$^{2}$
J. Westcott$^{7}$}
\\
$^{1}$School of Physics and Astronomy, University of Southampton, Southampton, SO17 1BJ, UK\\
$^{2}$Jodrell Bank Centre for Astrophysics, School of Physics and Astronomy, The University of Manchester, Manchester, M13 9PL, UK\\
$^{3}$University of Central Lancashire, Jeremiah Horrocks Institute Preston, UK PR1 2HE\\
$^{4}$Departamento de  Astrof\'isica y Ciencias de la Atm\'osfera, Universidad Complutense de Madrid, E-28040 Madrid, Spain\\
$^{5}$Instituto de Astrof\'{i}sica de Canarias, V\'{i}a L\'{a}ctea S/N, E-38205 La Laguna, Spain\\
$^{6}$Departamento de Astrof\'{i}sica, Universidad de La Laguna, E-38206 La Laguna, Spain\\
$^{7}$Centre for Astrophysics Research, University of Hertfordshire, College Lane, Hatfield, AL10 9AB, UK\\
$^{8}$Department of Physics \& Astronomy, University College London, Gower Street, London, WC1E 6BT, UK\\
$^{9}$University of Bath, Claverton Down, Bath, BA2 7AY, UK\\
$^{10}$INAF - IAPS Rome, Via Fosso del Cavaliere 100, I-00133 Roma, Italy\\
}

\date{Accepted 2017 August 21. Received 2017 August 21; in original form 2017 July 24}

\pubyear{2017}

\begin{document}
\label{firstpage}
\pagerange{\pageref{firstpage}--\pageref{lastpage}}
\maketitle

\begin{abstract}

We present high-sensitivity eMERLIN radio images of the Seyfert galaxy NGC\,4151 at 1.5\,GHz. We compare the new eMERLIN images to those from archival MERLIN observations in 1993 to determine the change in jet morphology in the 22 years between observations. We report an increase by almost a factor of 2 in the peak flux density of the central core component, C4, thought to host the black hole, but a probable decrease in some other components, possibly due to adiabatic expansion. The core flux increase indicates an AGN which is currently active and feeding the jet. We detect no significant motion in 22 years between C4 and the component C3, which is unresolved in the eMERLIN image. We present a spectral index image made within the 512\,MHz band of the 1.5\,GHz observations. The spectrum of the core, C4, is flatter than that of other components further out in the jet. We use HST emission line images (H$\alpha$, [O~III] and [O~II]) to study the connection between the jet and the emission line region. Based on the changing emission line ratios away from the core and comparison with the eMERLIN radio jet, we conclude that photoionisation from the central AGN is responsible for the observed emission line properties further than 4$\arcsec$ (360\,pc) from the core, C4. Within this region, several evidences (radio-line co-spatiality, low [O~III]/H$\alpha$ and estimated fast shocks) suggest additional ionisation from the jet.

\end{abstract}


%%%%%%%%%%%%%%%%%%%%%%%%%%%%%%%%%%%%%%%%%%%%%%%%%%

%%%%%%%%%%%%%%%%% BODY OF PAPER %%%%%%%%%%%%%%%%%%

\section{Introduction}
At the centre of every galaxy is thought to lie a super-massive black hole (SMBH) \citep{Magorrian98,Gebhardt,FerrareseMerritt00}. Broad-band emission from SMBHs is observed from the X-ray through to the radio regime. When they accrete matter they turn into active galactic nuclei (AGN) \citep{HoReview}. Most of the AGN in the local Universe are radio-quiet, defined by \cite{Terashima} as those where the logarithm of the ratio of the radio (5\,GHz) to X-ray (2$-$10\,keV) luminosity, denoted log(R$_X$), is $\ge$ $-$4.5. Kilo-parsec radio jets are seen in such AGN \citep{Condon87,Ulvestad2003,Ghisellini2004}. Radio variability has been detected 
in the nuclei of some such AGN, e.g. \cite{Wrobel99,Mundell2009} though not in others, e.g. \cite{SadieJones2011,SadieJones2017}. However temporal studies of larger scale jets in radio-quiet AGN are rare due to their intrinsic radio weakness. Thus, changes in jet morphology which might indicate jet motion or regions of particle acceleration cannot be measured. One exception is the jet in the well-known Seyfert 1.5 NGC\,4151. It is one of the brightest AGN in the sky in X-rays \citep{Gursky1971,Boksenberg95,Ogle00,Wang2010,Wang11} and the radio-brightest of the radio-quiet AGN \citep{4151BrightestRadioQuietAGN}. As such it is a great probe of the intermediate regime of radio-loudness, to explore the mechanisms of jet propagation through the interstellar medium (ISM). NGC\,4151 is a nearly face-on (\textit{i} $\approx$ 21$^{\circ}$) barred spiral galaxy. It has one of the most precise distance measurements of an AGN to date, due to dust-parallax measurements, of 19\,Mpc \citep{Honig14}. This corresponds to an angular scale of $\sim$91\,pc per arc second.% and a black hole mass of 5.4 $\times$ 10$^7$M$_{\odot}$ \citep{Onken14}, assuming this distance. 

\begin{figure}
	\includegraphics[width=0.5\textwidth]{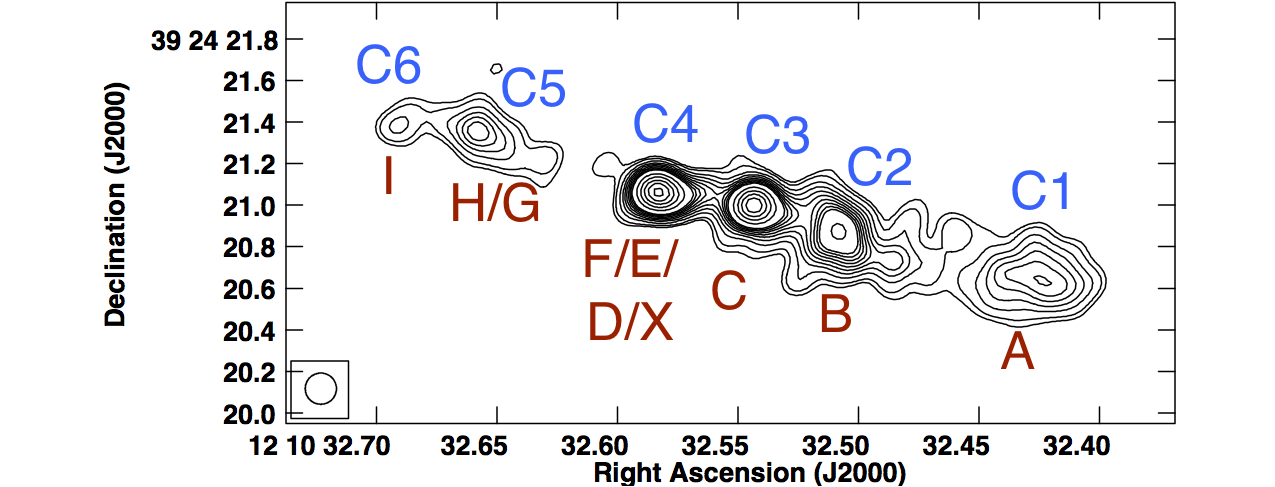}
    \caption{Naturally weighted archival MERLIN image of the central, 4 $\times$ 2$\arcsec$ ($\sim$360 $\times$ 180\,pc) radio structures of NGC\,4151, re-reduced with the MERLIN pipeline. The FWHM of the restoring beam was set to 0.15$\arcsec$ $\times$ 0.15$\arcsec$ (14 $\times$ 14\,pc) and the entire \textit{uv}-range with all 8 antennas in the MERLIN array was used to produce this image in {\scriptsize AIPS}. For consistency, the contours are the same as Fig.~2 in \citep{Mundell95}: 1.5, 2, 3, 4, 5, 6, 7, 8, 9, 10, 15, 20, 25, 30~mJy/beam. The naming conventions are shown above with \citet{Carral90} nomenclature in blue, and the \citet{Ulvestad2005} nomenclature in red.}
    \label{fig:OLDMERLINMAP}
\end{figure}

NGC\,4151 has been extensively studied in the radio for several decades \citep{WilsonUlvestad82,Johnston82,Booler82,Harrison86,Carral90,Pedlar93,Mundell95,Ulvestad98,Mundell03,Ulvestad2005}. The radio structure is characterised by a double-sided jet at PA $\sim$77$^{\circ}$ extending from a nucleus with VLBI centre at $\alpha_{J2000}$=12$^h$10$^m$32.5758$^m$ and $\delta_{J2000}$=+39$^d$24$^m$21.060$^s$ \citep{Ulvestad2005}. Archival VLA/MERLIN observations \citep{Carral90, Pedlar93, Mundell95} detect six radio components along this structure named C1 to C6 with the naming convention going from west to east. Further resolved components were discovered with VLBA/VLBI images \citet{Ulvestad2005}, who renamed the components A to I (Fig.~\ref{fig:OLDMERLINMAP}). Throughout this paper we shall use the original C1-C6 nomenclature, unless specified otherwise. The core radio component (named C4) is co-incident with the optical nucleus in \citet{Mundell95} (hereafter referred to as M95) leading to its identification as the AGN. Here we present deep 1.51 GHz observations with the upgraded eMERLIN radio interferometer allowing, by comparison with M95, study of changes in the jet morphology over a 22 year period.

%\begin{table}
%	\centering
%	\caption{Key properties of the AGN in NGC\,4151.}
%	\label{tab:example_table}
%	\begin{tabular}{lcp{3cm}r} % four columns, alignment for each
%		\hline\hline
%		Property & Value & Reference\\
%		\hline
%		$\alpha_{J2000}$ & 12$^h$10$^m$32.5758$^m$ & \citep{Ulvestad2005}\\
%		$\delta_{J2000}$ & +39$^d$24$^m$21.060$^s$ & \citep{Ulvestad2005}\\
%		Distance & 19.0\,Mpc & \citep{Honig14}\\
%		Black Hole Mass & 5.4 $\times$ 10$^7$M$_{\odot}$ & \citep{Onken14,Honig14}\\
%		AGN Type & Seyfert 1.5 & \citep{OsterbrockKoski76}\\
%		\hline
%	\end{tabular}
%\end{table}

With the exception of the decommissioning of the Wardle (Mk III) antenna, the configuration of eMERLIN is identical to that of MERLIN, providing an angular resolution of 150mas at 1.5 GHz. The bandwidth of eMERLIN is wider that that of MERLIN, leading to improved \textit{uv}-coverage. The eMERLIN observations of NGC\,4151 were made as part of the \textbf{L}egacy \textbf{e}-\textbf{M}ERLIN \textbf{M}ulti-band \textbf{I}maging of \textbf{N}earby \textbf{G}alaxies \textbf{S}urvey - LeMMINGs \citep{BeswickLemmings}. LeMMINGS is the second largest of the eMERLIN legacy surveys and consists of observations of all 280 galaxies above $\delta \ge 20^{\circ}$ from the Palomar sample of nearby galaxies \citep{Filippenko85,Ho95,Ho97a,Ho97b,Ho97e,Ho97c,Ho97d,Ho03,Ho09}.

A number of observers \citep{Perez1989,Evans1993,Robinson1994,Boksenberg95,Winge97,Hutchings98,Winge99,Hutchings99,Kaiser2000,Kramer2008} have shown optical emission line images of the nuclear regions of NGC\,4151. Whilst it is generally agreed that photoionisation from the AGN is an important contributor to the ionisation, the importance of the radio jet is not so clear. Here, by combining the eMERLIN image with \textit{HST} H$\alpha$, [O~II], and [O~III] images, we explore the contribution of the jet in more detail.

In section \ref{sec:ObservationsAndDataReduction} we present the eMERLIN observations and data reduction. In section \ref{sec:Results} we discuss morphological changes between the present image and the previous MERLIN image. In section \ref{sec:EmissionLines} we discuss the relationship between the radio jet and the optical line emission region and we summarise our conclusions in section \ref{sec:SummaryandConclusions}.

\section{Observations and Data Reduction}
\label{sec:ObservationsAndDataReduction}

\begin{figure}
	\includegraphics[width=\columnwidth]{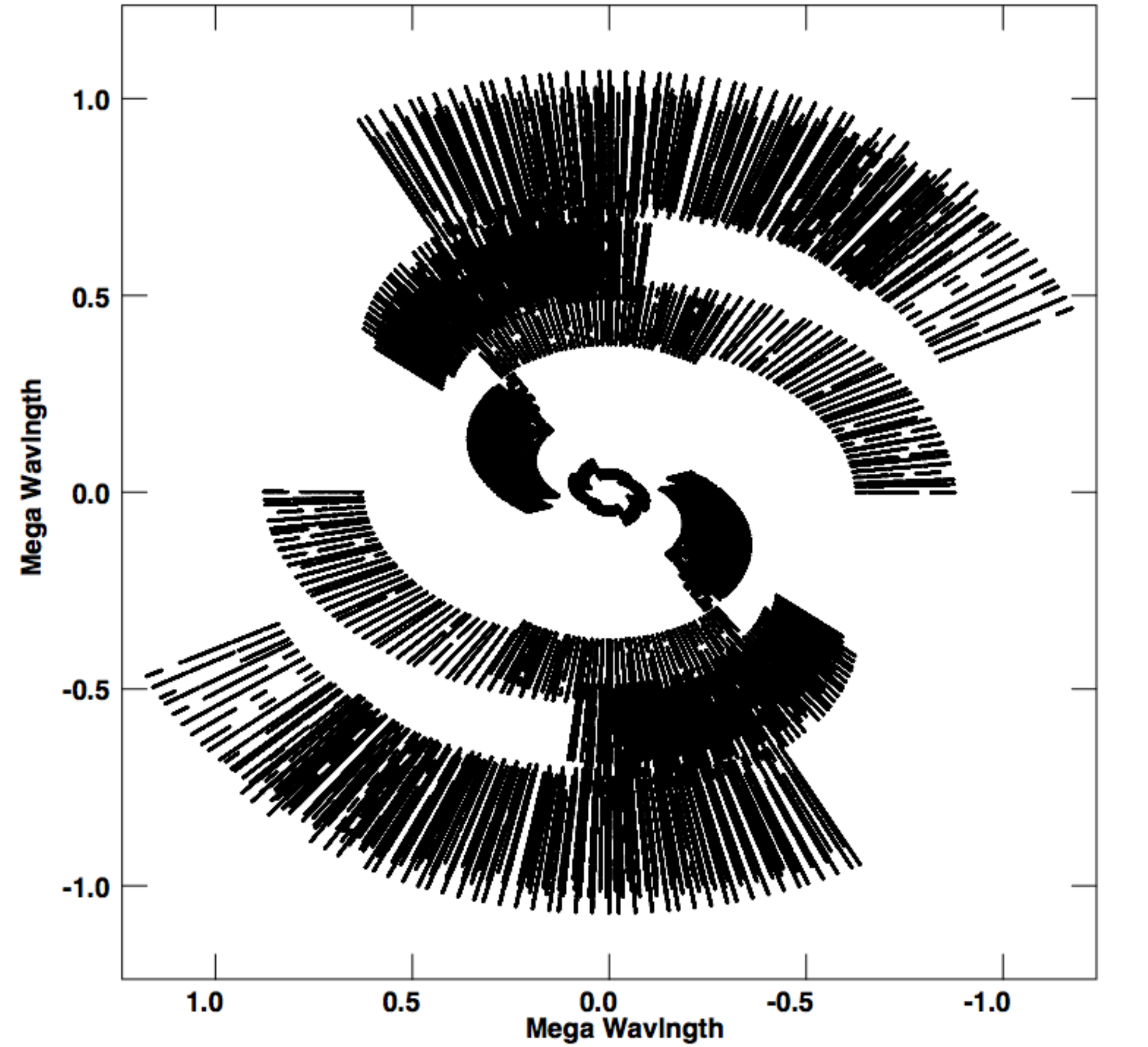}
    \caption{Full observed \textit{uv}-plane of NGC\,4151 at 1.5 GHz, using the LeMMINGs (eMERLIN) deep data with all 7 antennas included.}
    \label{fig:4151UVPLT}
\end{figure}

\subsection{eMERLIN Data Reduction}
\label{sec:Datareduction} % used for referring to this section from elsewhere

Observations of NGC\,4151 were performed at L-band (weighted central frequency of 1.51 GHz) with the eMERLIN array as part of the LeMMINGs deep sample. This sub-sample consists of a small number of galaxies for which particularly deep pbservations have been taken including NGC\,4151 (this paper), IC\,10 \citep{Westcott2017}, NGC\,5322 (Dullo et al in prep.) and NGC\,6217 (Williams et al in prep). All 7 antennas in the array participated in the observation on 29th April 2015, including the Lovell telescope. NGC\,4151 was observed on-source for 3.81 hours with data reduction and imaging following the steps outlined in the e-MERLIN cookbook and pipeline \citep{eMERLINcookbook,eMERLINpipeline} with {\scriptsize AIPS} \citep{AIPS}. The full \textit{uv}-plane of the eMERLIN observations is shown in Fig.~\ref{fig:4151UVPLT}. The observational set-up used a total bandwidth of 512\,MHz, centred on 1.51\,GHz. The 512\,MHz band was split into 8 intermediate frequencies (IFs) of width 64\,MHz and consisting of 128 channels in each IF. The calibrator NVSSJ120922+411941 (J1209+4119) was used for phase referencing and OQ208 and 3C286 were used as the band pass and flux calibrators respectively. The target and phase calibrator alternated during the observing run, with blocks of  approximately 2.5 mins on the phase calibrator and 7 mins on the target, with the flux and band pass calibrators observed at the end of the observing run. 

To calibrate the data, we followed the procedure outlined in the e-MERLIN cookbook \citep{eMERLINcookbook}, a summary of which we include below. Correlation and averaging of the data was performed before the {\scriptsize SERPent}\footnote{The {\scriptsize SERPent} flagging code \citep{SERPent} is written in {\scriptsize ParselTongue} \citep{ParselTongue}} flagging code, was used to remove the worst instances of radio frequency interference (RFI) from the data. The raw data was then inspected with {\scriptsize AIPS} tasks {\scriptsize SPFLG} and {\scriptsize IBLED} to remove any low-level RFI not picked up by the automatic flagger. In addition to the RFI flagging, the first two IFs of the LL polarisation were flagged on all Lovell baselines due to the inclusion of a test filter on the antenna. The channels showing no coherent phase at the ends of each IF were also flagged. It is estimated that approximately 15 per cent of the on-source data were flagged during this process and further calibration rounds before the final images were made.

To begin the calibration procedures, we fitted the offsets in delays using the {\scriptsize AIPS} task {\scriptsize FRING} before calibrating for phase and gain. Band pass solutions were also calculated with {\scriptsize BPASS} and applied followed by imaging of the phase calibrator, which was self-calibrated until solutions converged. The complex antenna solutions from self-calibration of the phase calibrator were then applied to the target field.

\begin{figure*}

	\includegraphics[width=0.9\textwidth]{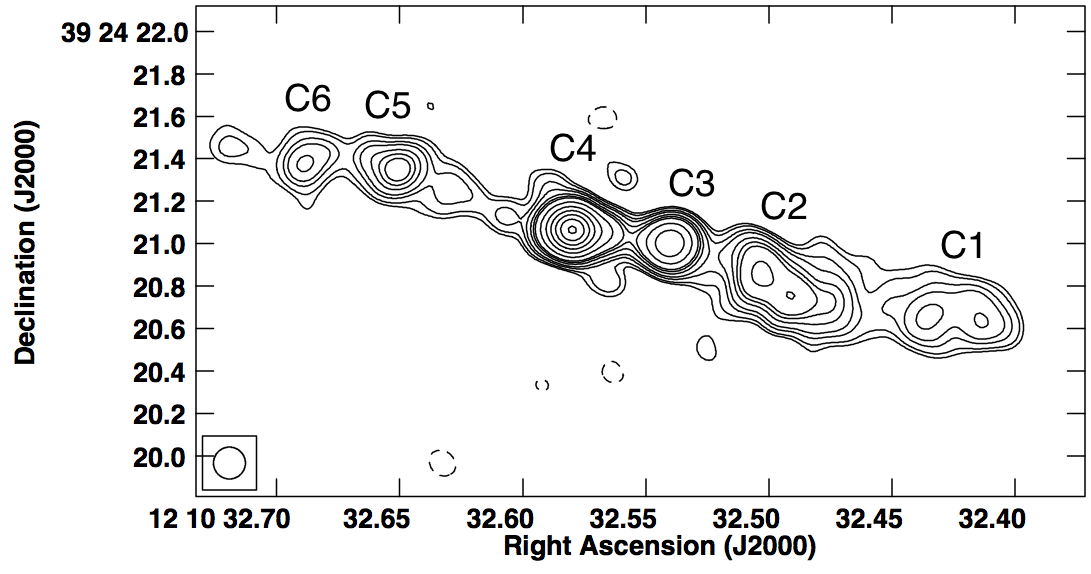}
    \caption{New full-resolution eMERLIN image of the central 4 $\times$ 2$\arcsec$ ($\sim$360 $\times$ 180\,pc) region of NGC\,4151 using all 7 eMERLIN antennas and a natural weighting. As in Fig.~\ref{fig:OLDMERLINMAP}, the entire \textit{uv}-range was used with a 0.15$\arcsec$ $\times$ 0.15$\arcsec$ FWHM restoring beam. Contours set are at -0.25, 0.75, 1, 1.5, 2, 3, 4, 5, 9, 16, 25, 36, 49, 64 mJy/beam.}
    \label{fig:FULLRESMAPDOTV1}

\end{figure*}

\subsubsection{eMERLIN Radio Imaging}
\label{sec:Imaging} % used for referring to this section from elsewhere
These data were imaged with {\scriptsize IMAGR} and phase self-calibration was applied to further refine the data before the visibilities were re-weighted using {\scriptsize REWAY} to account for variable sensitivity as a function of antenna, as  eMERLIN comprises of an inhomogeneous set of antenna types, frequency and time to maximise the resultant sensitivity of the data. Further self calibration improved the signal-to-noise and the final image was created with the noise in the naturally weighted image of 35$\mu$Jy as shown in Fig.~\ref{fig:FULLRESMAPDOTV1}. Due to the complex nature of this source and its brightness, great care was taken to only include real what are deemed to be genuine emission features in the self calibration process. An amplitude and phase self calibration did not produce stable phases or amplitudes, so only the final phase self calibrated data are shown. This data set was then used to create all further images of NGC\,4151. 

\subsubsection{Archival MERLIN Data Reduction and Imaging}
Archival data originally published in M95 were re-reduced to compute accurate positions and flux density measurements to compare to the new eMERLIN data set. NGC\,4151 was observed in November 1993 at 1.42\,GHz. The MERLIN data was calibrated with the MERLIN pipeline\footnote{www.merlin.ac.uk/archive/} in {\scriptsize AIPS} and re-imaged to account for RFI, alter the beam size and re-weight the data. As the data were originally taken in spectral line imaging mode to study the neutral hydrogen absorption in NGC\,4151 (see M95), to create a continuum image we flagged all channels that were contaminated by this line to ensure the flux measurements on the core component C4 were not affected by this absorption feature. The amount of on-source time was 9.8 hours, achieving an rms noise of 0.25\,mJy/beam. 
We thus see the large improvement in S/N achieved by eMERLIN compared to the previous MERLIN array. All stations in the MERLIN array took part in this observation including the Lovell telescope. All images produced were made at the same resolution as those published by M95, in order to be able to directly compare all data. The re-reduced natural image of the data is shown in Fig.~\ref{fig:OLDMERLINMAP} together with the contouring scheme from M95. Note that the extended low surface brightness structures seen to the south of C2 in Fig.2 of M95 is not present in our re-reduced image, probably due to improved imaging fidelity of these new eMERLIN data due to the large bandwidth (512\,MHz compared to 8\,MHz) resulting in more complete sampling of spatial scales.

\subsubsection{Comparison between the two radio epochs}
When comparing the two epochs of data, some caution needs to be taken since archival MERLIN data does not completely match the new eMERLIN data in terms of \textit{uv}-range, number of antennas, bandwidth, central frequency and observing time. 

Therefore, to make an unbiased comparison of the source structure at the two data epochs, we made two additional images with similar conditions for each dataset. The \textit{uv}-range was limited to between 100 and 1000 k$\lambda$ where the two datasets overlapped. We exclude the now defunct Mark III (Wardle) antenna from the MERLIN data. All of the images were made with the same FWHM restoring beam size of 0.15$\times$0.15$\arcsec^{2}$, corresponding to 14 $\times$ 14\,pc. The final images are shown in Fig.~\ref{fig:NEWMERLINMAP-AN4401000UV} and
Fig.~\ref{fig:EMERLINMAP401000UV} based on MERLIN and eMERLIN respectively. Subsequently, this enables us to compare emission on the same spatial scales. 

The eMERLIN data were then loaded into {\scriptsize CASA} to produce an in-band spectral index image with {\scriptsize clean} and {\scriptsize nterms} set to 2. The spectral index image is shown in Fig.~\ref{fig:CASASpectralIndexMap}.

\subsection{Optical Data}

\cite{Perez1989} presented [O~III] and H$\alpha$ ground-based imaging of NGC\,4151 and show that the ratio of [O~III]/H$\alpha$ defines a large ($\sim$10$\arcsec$, 910\,pc) cone-like structure to the south-west of the nucleus of PA $\sim$50$^{\circ}$, misaligned with respect to the radio structure. \cite{Evans1993} and \cite{Boksenberg95} have presented \textit{HST} Planetary Camera (PC) images in [O~III] and H$\alpha$. HST Wide Field Planetary Camera 2 (WFPC2) images in [O~II], [O~III] and H$\alpha$ have been presented by \citep{Hutchings98,Hutchings99,Kaiser2000,Kramer2008}. 

High-resolution {\it HST} Wide-Field Planetary Camera 2 (WFPC2) images of NGC\,4151 were taken, from the GO-5124 program (PI: H. Ford), in the F336W, F375N ([O~II]), F502N ([O~III]), F547M, F658N (H$\alpha$) and F791W filters were retrieved from the public Hubble Legacy Archive (HLA\footnote{http://hla.stsci.edu}). The data were observed on the 22nd January 1995. Here we re-reduce these data, using the state-of-the-art techniques described in \cite{DulloMartinezLombilla2016},
providing [O~II], [O~III] and H$\alpha$ emission line images and produce an \textit{HST} [O~III]/H$\alpha$ image which has not been previously presented. These images are considered, together with the eMERLIN radio image, in Section~\ref{sec:EmissionLines}.

In order the create the emission line images we used the WFPC2 PC1 F336W, F375N, F502N, F547M and F658N images and followed the procedures outlined in \cite{KnapenStedman2004}, \cite{SanchezGallegoKnapen2012} and \cite{DulloMartinezLombilla2016} and compared the narrow-band F375N ([O~II]) image with the broad-band F336W image to create the [O~II] continuum-subtracted emission line image (Fig.~\ref{fig:O2optical}). Similarly, we created the [O~III] and H$\alpha$ emission line images by comparing the F502N ([O~III]) and F658N (H$\alpha$) images with the F547M image of the galaxy. Figs.~\ref{fig:O2optical} and ~\ref{fig:HaOIIsub} show our new eMERLIN L-band image overlaid on these three ([O~II], [O~III] and H$\alpha$) line images and on the [O~III]/H$\alpha$ emission line ratio, respectively. 

The $V-H$ dust map of NGC\,4151 by \citet[their Fig.~1]{Martini2003} shows that the galaxy has spiral dust arms at $R \ga 100$ pc. However, this dust map and the visual inspections of the {\it HST} images (see Fig.~\ref{fig:O2optical}) show that the emission-line regions in the galaxy are only weakly obscured by dust. Therefore, we did not attempt to correct for the spiral dust arms, although dust extinction may somewhat affect our [O~II] line flux measurement since dust absorption is relatively higher at shorter wavelengths. However, when we extract emission line fluxes from the images (Section~\ref{sec:EmissionLines}), we take into account the reddening using the extinction law from \citet{CalzettiLaw}.

We note that the coordinates of the radio core (component C4) in the eMERLIN image (Fig.~\ref{fig:FULLRESMAPDOTV1}) differ by 0$\farcs$2 (18\,pc) from those of the continuum core in the {\it HST} images. The radio positions are linked to the positions of VLBI phase calibrator sources. Systematic positional uncertainties are $\la$ 1 mas, which can be neglected here. The {\it HST} positions are linked to the positions of stars in the Guide Star Catalogue \citep{GuideStarCatalogue} which are typically accurate to 0$\farcs$3 (27\,pc at this distance). Thus the discrepancy between the radio and optical core positions is likely attributed to uncertainties in the {\it HST} positions. We therefore align the radio and optical images so that the maximum surface brightness at the core of each optical image is coincident with the radio core, C4. We only move the images in Right Ascension and Declination and do not rotate. To align the radio and optical data we used {\scriptsize astropy\footnote{http://docs.astropy.org/}} to edit the FITS file header for the central Right Ascension and Declination values.

\section{Radio Morphology and Spectral Index}
\label{sec:Results}

\begin{table*}
	\centering
	\caption{Flux densities obtained from Figs.~\ref{fig:NEWMERLINMAP-AN4401000UV} and \ref{fig:EMERLINMAP401000UV}, where the resolution of the data is matched, of each component in NGC\,4151. The spectral index was obtained from the spectral index image in Fig.~\ref{fig:CASASpectralIndexMap}, the size of the components as found from the fitting process in section \ref{sec:Results} and the minimum energy and magnetic field obtained from this process. 
}
	\label{tab:FLUXTABLE}
	\begin{tabular}{lp{1.5cm}p{2.0cm}p{1.5cm}p{2.0cm}p{1.5cm}p{2.5cm}p{1.0cm}p{1.0cm}r} % four columns, alignment for each
		\hline\hline
		Comp. & Peak Flux Density MERLIN (mJy\,beam$^{-1}$) & Integrated Flux Density MERLIN 1993 (mJy) & Peak Flux Density eMERLIN (mJy\,beam$^{-1}$) & Integrated Flux Density eMERLIN (mJy) & Spectral Index eMERLIN & 	Component Size in eMERLIN image (mas$^2$)& B$_{min}$ (mG) & log(U$_{min}$) (J) \\
		\hline
		C1 & 5.92 $\pm$0.06 & 2.74 $\pm$0.35 & 1.94 $\pm$0.04 & 6.09 $\pm$0.05 & -0.9 & 346.8 $\times$ 203.4  & 0.39 & 44.02\\
		C2 & 13.64 $\pm$0.07 & 28.69 $\pm$0.20 & 6.51 $\pm$0.04 & 15.36 $\pm$0.04 & -0.9 & 327.0 $\times$ 162.9 & 0.59 & 44.16\\
		C3 & 31.66 $\pm$0.07 & 42.38 $\pm$0.15 & 22.92 $\pm$0.04 & 25.74 $\pm$0.02 & -0.7 & 172.0 $\times$ 147.0 & 0.76 & 44.01\\
		C4 & 37.14 $\pm$0.07 & 46.68 $\pm$0.14 & 66.76 $\pm$0.04 & 72.00 $\pm$0.02 & -0.4 & 161.2 $\times$ 150.6 & 0.96 & 44.21\\
		C5 & 6.05 $\pm$0.07 & 10.69 $\pm$0.17 & 4.83 $\pm$0.04 & 6.85 $\pm$0.03 & -0.9 & 202.8 $\times$ 157.3 & 0.54 & 43.85\\
		C6 & 1.87 $\pm$0.07 & 1.98 $\pm$0.13 & 1.95 $\pm$0.04 & 1.86 $\pm$0.02 & -0.9 & 156.5 $\times$ 137.5 & 0.44 & 43.43\\
		\hline
	\end{tabular}

\end{table*}

The full-resolution naturally-weighted 1.5\,GHz eMERLIN image of NGC\,4151 (Fig.~3) shows the six previously known  components. C1-C6, extending over 4$\arcsec$ (360\,pc), similar to the twin-jet morphology observed in M95. The morphology shown in this image is similar to the naturally weighted MERLIN image (Fig.\ref{fig:OLDMERLINMAP}) but the core, C4, is definitely brighter, by a factor of 1.5, in the eMERLIN image. The jets, particularly the western jet, also appear narrower. 

\begin{figure*}
	\includegraphics[width=0.9\textwidth]{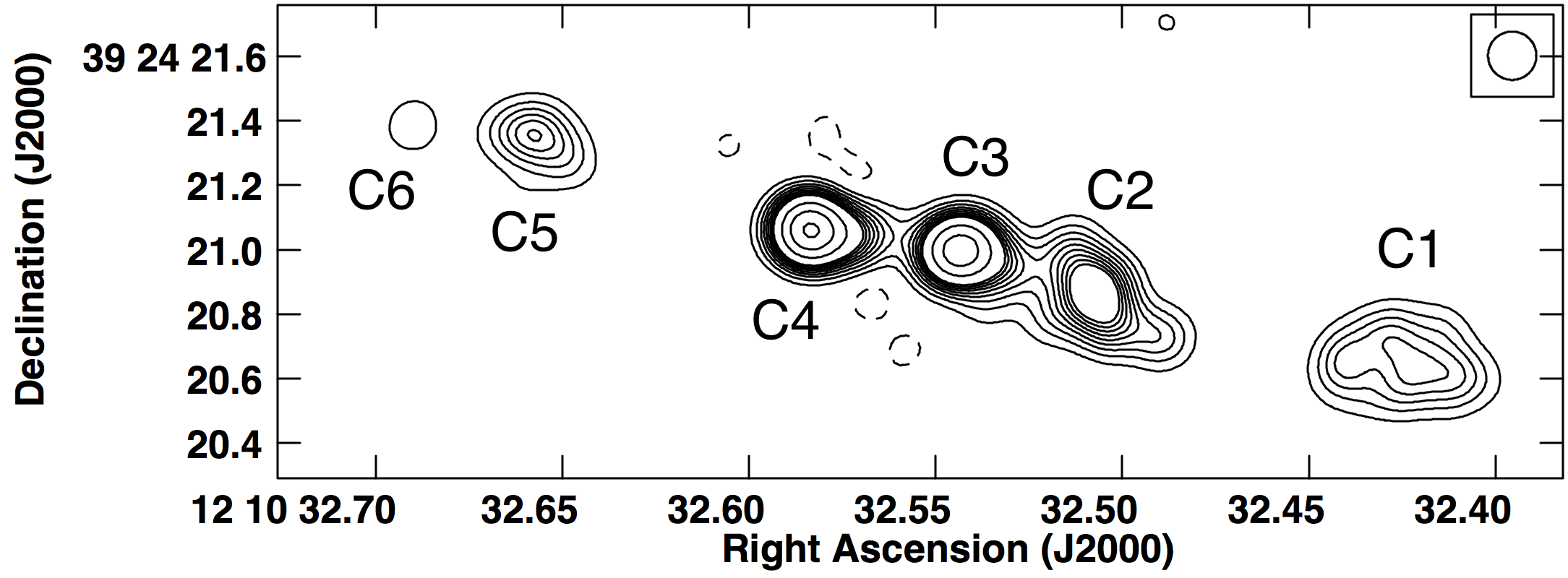}
    \caption{Naturally weighted archival MERLIN image of the central 3.6 $\times$ 1.2$\arcsec$ ($\sim$330 $\times$ 110\,pc) region of NGC\,4151, re-reduced with the MERLIN pipeline here. The beam was set to 0.15$\arcsec$ $\times$ 0.15$\arcsec$ and a restricted \textit{uv}-range of 100 $-$ 1000 k$\lambda$ was used to produce this image in {\scriptsize AIPS} to overlap with the same \textit{uv}-range in the eMERLIN image. All antennas except the Mk~III (Wardle) in the MERLIN array were used so as to include only those antennas in the current eMERLIN array. The contours are -0.75, 1, 2, 3, 4, 5, 6, 7, 8, 9, 16, 25, 36, 49, 64 mJy/beam. The naming convention from \citet{Carral90} are overlaid in black.}
    \label{fig:NEWMERLINMAP-AN4401000UV}
    	\includegraphics[width=0.9\textwidth]{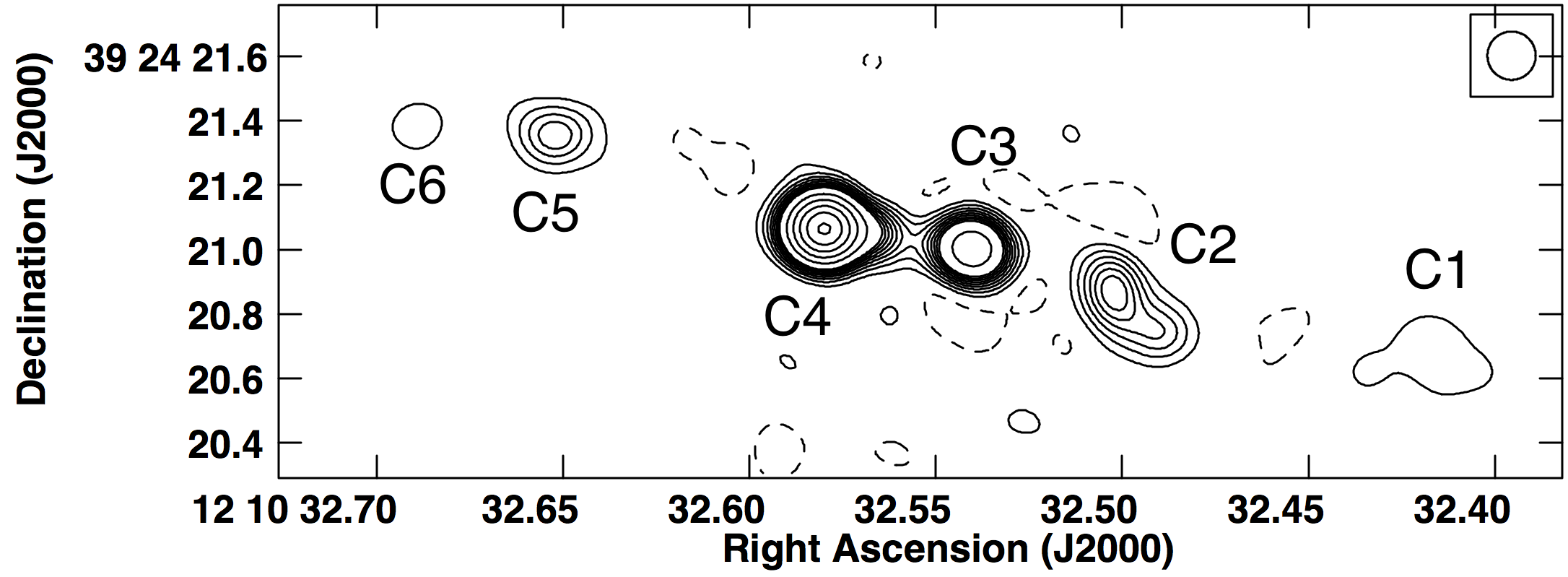}
    \caption{Same as Fig.~\ref{fig:NEWMERLINMAP-AN4401000UV} but using the new eMERLIN data and reduced with the eMERLIN cookbook. This included all 7 eMERLIN antennas. As in Fig.~\ref{fig:NEWMERLINMAP-AN4401000UV}, the \textit{uv}-range was limited to 100 - 1000 k$\lambda$.}
    \label{fig:EMERLINMAP401000UV}
\end{figure*}

Below we will consider the morphological changes in the radio components, from  C1 to C6, between the MERLIN and eMERLIN images, as well as changes in positions and fluxes. We fitted 2D Gaussian components to each of the components C1-6 using the {\scriptsize AIPS} task {\scriptsize JMFIT}. The peak flux density, integrated flux, RA and Dec, and size of each component were extracted and are shown in Tables \ref{tab:FLUXTABLE} and \ref{tab:ASTROMETRYTABLE}. The positions of the components stated in Table~\ref{tab:ASTROMETRYTABLE} refer to the self-calibrated data. These positions are consistent with the values from the dirty image without any self-calibration applied, to within 0.001$\arcsec$, much smaller than the FWHM of the restoring beam. Note that in those tables we list only the error given by {\scriptsize JMFIT} from fitting the components. The true error, including the rms noise level on the image, possible contamination by sidelobes from other components and uncertainties in the flux density calibrator, is hard to define but is likely to be at least 3 times larger. In the following sections we will focus on each component of the jet in turn to study their morphology, possible movement and flux density change with respect to the two epochs. In terms of the radio morphology, the full resolution images are compared, while flux density variations come from the comparison of the \textit{uv}-range restricted images (Fig.~\ref{fig:NEWMERLINMAP-AN4401000UV} and Fig.~\ref{fig:EMERLINMAP401000UV}).

We caution, however, that small differences in \textit{uv}-coverage can have a noticeable effect in imaging, particularly of low brightness structures. Here, although we are able to restrict the 1D \textit{uv}-range to be the same for the MERLIN and eMERLIN images (Figs.~\ref{fig:NEWMERLINMAP-AN4401000UV} and ~\ref{fig:EMERLINMAP401000UV}), we are not able to take the final step of restricting the 2D \textit{uv}-coverage of the MERLIN image to be the same as for eMERLIN as the reduction in sensitivity is then too great to allow useful imaging of low brightness structures.

\subsection{ The Individual Components}
\label{sec:Morphology}

\subsubsection{C1 and C2}
Comparing Figs.~\ref{fig:NEWMERLINMAP-AN4401000UV} and ~\ref{fig:EMERLINMAP401000UV}, the overall peak flux density of C1 and C2 appears to have decreased. However we caution that the different \textit{uv}-coverage of the MERLIN and eMERLIN datasets may affect the detection of large scale structure, giving rise to negative bowls underlying the more compact emission and hence reducing peak flux densities, in the eMERLIN image. In this respect it is interesting to note that with the lower contour levels for the full-resolution eMERLIN image (Fig.~\ref{fig:FULLRESMAPDOTV1}), the extended structures of C1 and C2 resemble those of the MERLIN image (Fig.~\ref{fig:OLDMERLINMAP}).

\subsubsection{C3 and C4}
C3 is unresolved, even at the highest angular
resolution achievable with eMERLIN. Similar to components C1 and C2, the flux density in C3 decreased by a factor of $\sim$1.5. We again caution that at least part of this change may be due to changes in 2D \textit{uv}-coverage but for an unresolved component the differences in \textit{uv}-coverage should not be so important. Higher-resolution 1.4\,GHz VLBA observations
\citep{Ulvestad98} show a faint (2.6mJy total flux density) component
at the location of C3. This indicates that the emission associated
with component C3 by MERLIN and eMERLIN must be related to an extended
and diffuse region, undetectable with the VLBA. 

The core, where the jet base is located, corresponds to component
C4. It is slightly elongated on the western side towards C3,
where the 4 mJy/beam  contour level connects it to C3 as it did in the
previous M95 image. In addition, other weaker extensions are associated
with C4 towards the north, south and east. We cannot conclude anything
about their nature because of their small size and weakness. 

C4 is the brightest component, with a peak flux
density of $\sim$67 mJy/beam, higher than in the MERLIN observations by
nearly a factor of $\sim$2 and corresponding to a luminosity of $3.87 \times 10^{37}$\,ergs\,s$^{-1}$.
VLBI observation at 18\,cm \citep{Ulvestad98,Mundell03,Ulvestad2005} with an angular resolution of
a few mas show several components along the jet axis located within the
core C4, which are not resolved by eMERLIN. The flux density of one of these VLBI components (D3b) increased by $\sim$30 per cent in 4 years between the observations of 
\citet{Ulvestad98} and \citet{Ulvestad2005}. The larger flux density changes seen between the MERLIN and eMERLIN observations are quite consistent with these VLBI changes and confirm that C4 contains a currently active AGN core, injecting relativistic particles into the inner jet.

\subsubsection{C5 and C6}
Moving to the eastern side of the jet, a 0.4$\arcsec$ (36\,pc) significant elongation bridges C4 and C5 (Fig.~\ref{fig:FULLRESMAPDOTV1}). C5 has a clear extension to the west,
similar to that in the 1993 observations. Furthermore, this extension overlaps
with a VLBA component G \citep{Mundell03}. The peak intensity of component C5 remains mostly unchanged. C6 is the last component of the
eastern jet and is consistent in flux and morphology with previous
MERLIN images. This element has a PA 125$^\circ$, clearly different from
the other components, suggesting a possible bending of the jet
spine. A further weak component appears
east of C6 which is not present in the previous MERLIN data. This component requires further investigation with eMERLIN to confirm its presence and morphology.

\subsubsection{Summary}
C4 has definitely increased in flux density (by factor nearly 2) in the 22 year period between the MERLIN and eMERLIN observations. Components C5 and C6 in the eastern jet have not changed noticeably. This lack of change leads confidence to the finding of a decrease in flux density of components C3, C2 and C1 in the western jet is real.

\begin{table*}
	\centering
	\caption{Astrometric measurements of radio components compared to the position of the core C4 obtained from the self-calibrated data. As the data are self-calibrated, we caution that the absolute values of RA and Dec are not absolute. These positions are compared to observations first published in M95, but have been re-reduced here (see section \ref{sec:Datareduction}). All positions given are in J2000 co-ordinates, with the difference calculated from the core C4 in that given image. The difference of these two values is shown in column 8 which is used as a measure of the relative shift of each component. The errors in the position from the fitting process are of the order half the beam size, i.e. $\sim$0.075$\arcsec$ (7\,pc).}
	\label{tab:ASTROMETRYTABLE}
	\begin{tabular}{lcccccccr} % four columns, alignment for each
		\hline\hline
		Comp. & RA eMERLIN & Dec eMERLIN & Difference & RA MERLIN & Dec MERLIN & Difference & Relative Shift\\
		\hline
		C1 & 12 10 32.41726 & +39 24 20.6601 & 1.928$\arcsec$ & 12 10 32.42493 & +39 24 20.6475 & 1.877$\arcsec$ & +0.051$\arcsec$ \\
		 %& $\pm$ 0.0000742s & $\pm$ 0.000505$\arcsec$ & & $\pm$ 0.0002224s & $\pm$ 0.001578$\arcsec$ & & \\
		C2 & 12 10 32.49967 & +39 24 20.8362 & 0.959$\arcsec$ & 12 10 32.50832 & +39 24 20.8694 & 0.886$\arcsec$ & +0.073$\arcsec$ \\
		 %& $\pm$ 0.0000163s & $\pm$ 0.000198$\arcsec$ & & $\pm$ 0.0000491s & $\pm$ 0.000631$\arcsec$ & & \\
		C3 & 12 10 32.54025 & +39 24 21.0026 & 0.466$\arcsec$ & 12 10 32.54319 & +39 24 20.9964 & 0.464$\arcsec$ & +0.002$\arcsec$ \\
		 %& $\pm$ 0.0000033s & $\pm$ 0.000034$\arcsec$ & & $\pm$ 0.0000181s & $\pm$ 0.000181$\arcsec$ & & \\
		C4 & 12 10 32.57989 & +39 24 21.0646 & - & 12 10 32.58279 & +39 24 21.0597 & - & - \\
		 %& $\pm$  0.0000011s & $\pm$ 0.000012$\arcsec$ & & $\pm$ 0.0000169s & $\pm$ 0.000161$\arcsec$ & & \\
		C5 & 12 10 32.65199 & +39 24 21.3570 & 0.884$\arcsec$ & 12 10 32.65756 & +39 24 21.3505 & 0.913$\arcsec$ & -0.029$\arcsec$ \\
		 %& $\pm$ 0.0000183s & $\pm$ 0.000167$\arcsec$ & & $\pm$ 0.0001366s & $\pm$ 0.001282$\arcsec$ & & \\
		C6 & 12 10 32.68911 & +39 24 21.3844 & 1.306$\arcsec$ & 12 10 32.68911 & +39 24 21.3885 & 1.275$\arcsec$ & +0.031$\arcsec$ \\
		 %& $\pm$ 0.0000355s & $\pm$ 0.000376$\arcsec$ & & $\pm$ 0.0002475s & $\pm$ 0.002773$\arcsec$ & & \\
		\hline
	\end{tabular}
	
\end{table*}

\subsection{Estimation of the jet speed}
\label{JetSpeed}
The intrinsic velocity, \textit{v}, of a component in a jet over a given time period is related to the angle that the line of sight subtends with the angle of the jet, \textit{$\theta$}, and its measured apparent transverse velocity \textit{v$_0$} for relativistic jets \citep{Kellerman1989}.

\begin{equation}
v= \frac{v_0}{[sin(\theta)+\frac{v_0}{c}cos(\theta)]}
\label{eq:velocity2}
\end{equation}

\cite{Pedlar93}, \cite{Robinson1994} and \cite{Vila-Vilaro1995} estimate that the pointing angle of the radio jet with respect to the line of sight is $\sim$40$^{\circ}$ based on a combination of geometric arguments, velocity images of the narrow line region \citep{Winge97,Hutchings98,Winge99} and the galactic disc inclination. We take this pointing angle into account when calculating the jet speed.

To accurately estimate the jet speed, components with reliable positions are needed. Hence only the unresolved components with no significant extensions can provide a precise astrometric position by fitting them with a 2D-Gaussian provided by the {\scriptsize JMFIT} task. This is the case for C3 and C4. As the data is self calibrated, we lose absolute astrometric positions of the components between the two epochs of data. However we do not lose relative positions and so we can still measure possible changes in separation between C3 and C4. We measure a change in separation of 2 mas but given a beam size of 150mas, we do not claim any detectable movement. Taking the 2mas figure at face value would give a velocity of 0.04c, completely consistent with the upper limits measured from VLBI observations by \cite{Ulvestad2005} but over a period of 4 years.  

\subsection{Radio Spectral Index}

\begin{figure*}
\centering
	\includegraphics[width=0.99\textwidth]{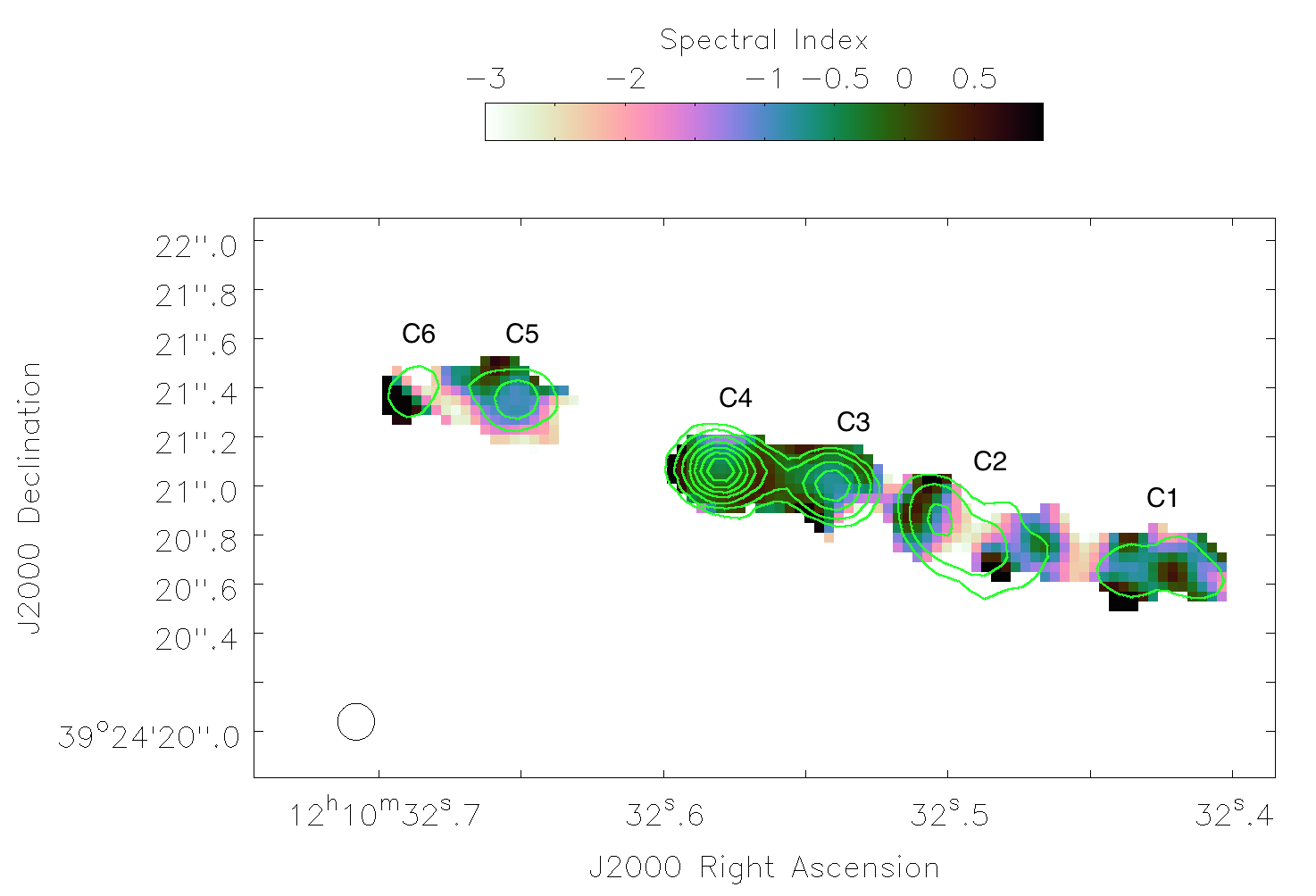}
    \caption{{\scriptsize CASA} eMERLIN spectral index image of the central 3.7 $\times$ 1.8$\arcsec$ ($\sim$340 $\times$ 160\,pc) of NGC\,4151 at 1.51\,GHz. The contours plotted are 2, 4, 9, 16, 25, 36, 49, 64 mJy/beam from the full resolution eMERLIN image. Data were clipped at 2mJy in the full resolution {\scriptsize CASA} eMERLIN image so that spurious low signal-to-noise regions were removed. All data were displayed with {\scriptsize CASA} and the range of colours reduced to $-$3.0 to 1.0 so as to remove any further spurious regions of non-physical spectral index from the image. The naming convention from \citet{Carral90} are overlaid in white.}
    \label{fig:CASASpectralIndexMap}
\end{figure*}

We use the definition of radio spectral index, $\alpha$, where  the flux density, $S$ at frequency $\nu$ is given by S$_{\nu}$ $\sim$ $\nu ^\alpha$.
A flat spectrum ($\alpha$ $\gtrsim$ $-$0.3) is
representative of a compact synchrotron self-absorbed source. Most AGN cores have flat spectra whereas extended, optically thin, emission has a steeper spectrum ($\alpha$ $\lesssim$ $-$0.7).

%As the electrons age, the spectrum is expected to steepen ($\alpha$ $\lesssim$ $-$0.5).

We derive the spectral index for the full-resolution eMERLIN image,
using the full 512\,MHz bandwidth. We use the {\scriptsize CASA} task {\scriptsize clean}
with {\scriptsize nterms} = 2 to interpolate through all the intermediate
frequencies \citep{RauCornwell}. We consider only the emission above
2mJy/beam to remove unreal fluctuations due to low S/N.
The spectral index image is presented in
Fig.~\ref{fig:CASASpectralIndexMap}. The calculated $\alpha$ values
range between $-$3 and 1.

Table~\ref{tab:FLUXTABLE} contains the radio spectral index of the components
along the jet, ascertained from Fig.~\ref{fig:CASASpectralIndexMap} at the location
of the flux peak of each component. The unresolved bright components, i.e. C3 and C4,
provide the most reliable spectral indices. The core C4 shows the flattest radio spectrum
within the eMERLIN bandwidth with a value of $\sim$ $-$0.4. This is
consistent with $\alpha$ obtained with wider-band
spectra by \cite{Pedlar93} (5$-$8\,GHz) and by \cite{Carral90} (1.6$-$15\,GHz),
both with the MERLIN array. The component C3 has a steeper index $\sim$ $-$0.7, consistent with the broad-band values in the literature
from MERLIN \citep{Carral90,Pedlar93} and also with our assertion that most of the flux density from C3 comes from a region which is extended on VLBI scales.
 
The remaining components show a steeper spectra still, although the lower S/N in these regions leads to large sporadic variations at the edges of components which are almost certainly not real.  Broadly, we note a steepening spectrum further away from the core, consistent with the classical picture of a jet losing energy by electron ageing and expansion \citep{Condon2016}. 

\subsection{Estimation of the magnetic field along the jet}
\label{MagField}

One possible explanation of a reduction in radio luminosity is simple radiative losses. To estimate the loss time-scale it is necessary first to estimate the magnetic field strength, which we do here.

Assuming the minimum energy condition applies to our target, we can
estimate the magnetic field $B_{min}$ of each component along the jet. This
assumption folds in information on the spectral index, size, flux
density, the jet composition and the pitch angle of the magnetic field. We
assume that the depth of the source is equivalent to the minor axis of
the component, unity in the ratio of energy in electrons and ions, no
angular dependence on the direction of the magnetic field and the line
of sight and a frequency range of 0.01$-$100\,GHz. The observational
constraints come from the integrated flux densities, the spectral
indices and the deconvolved sizes of the components shown in
Tab.~\ref{tab:FLUXTABLE}. 

We compute the magnetic field for all the components. However, as
discussed above, the low signal-to-noise ratio can significantly degrade the
accuracy of the fitting parameters for each component. Since C3 and C4
are bright resolved components, the estimates of their magnetic fields are likely more
reliable and are of the order of 1mG. The remaining components show slightly
smaller values of $B_{min}$, decreasing down the jet. \cite{Booler82}, performing a similar analysis
for NGC\,4151 using MERLIN data, estimated magnetic fields of the
same order.

VLBI/VLBA observations resolve further structure in components C3 and
C4, corresponding to smaller physical regions in the jet
\citep{Ulvestad98,Mundell03,Ulvestad2005}. Therefore, we calculate
$B_{min}$ using these higher-resolution data (flux,
size\footnote{Since the VLBI/VLBA observations do not provide the
  deconvolved sizes of the components, we used a component size half the size of the beam: 1.6 $\times$ 0.9 mas$^2$ for VLBI and
  2.9 $\times$ 2.4 mas$^2$ for VLBA.} and spectral index) for the VLBI
central core, D3, located at our component C4 and the VLBA component
C, located at C3. The $B_{min}$ for D3 is 5.38 $\times$
10$^{-2}$ Gauss, while it is 1.35 $\times$ 10$^{-2}$ Gauss for VLBA component C. We
note that these values are larger by a factor of several tens than the eMERLIN values due mainly to the smaller size of the component used when calculating $B_{min}$. Broadly, the range of $B_{min}$ values obtained with
this approach are consistent with the values of the magnetic field of radio
jets from the literature for AGN jets/cores \citep{Marscher1985, Biretta1991, Jester2005, WorrallBirkinshaw}.

\subsection{The cause of the reduction in radio flux density: synchrotron cooling or adiabatic losses?}

The synchrotron decay time scale, $\tau$, is related to the magnetic field strength 
and observing frequency by $\tau \propto$ B$^{-1.5}$ $\nu^{-0.5}$. Although much of the emission from C3 is extended on scales larger than those probed by VLBI, we can estimate the fastest decay time-scale by taking the magnetic field strength from VLBI measurements. This decay time-scale for C3 is $\sim$700 years. Thus in 22 years we would only expect a 3 per cent decrease, compared to the observed decrease of nearly 30 per cent. For the lower surface brightness components, C1 and C2, the time-scales would be longer. Thus, assuming that the flux density decrease is real, additional energy loss processes are required, e.g. adiabatic expansion losses. \citet{ScheuerWilliams1968} show that, for a linear expansion factor F (and for $S_{\nu} \propto \nu^{+\alpha}$), the flux density at a particular frequency will change by $F^{(4\alpha-2)}$. Thus for $\alpha$ $=$ $-$0.70 from our spectral fitting of component C3, the flux density changes by $\sim$ $F^{-5}$. Thus we require only a very small expansion (F=1.06) to produce a decrease of 25 per cent. The source size is not known but to explain the flux seen by eMERLIN but not detected by VLBA, will probably exceed a few VLBA beam sizes, i.e. $\sim$1\,pc. 

Given this size and expansion factor, the expansion speed of the component C3 in 22 years would be $\sim$2600\,km\,s$^{-1}$ but as the source size and decrease in flux are not accurately known, this speed is uncertain. However velocities of order $\sim$1000\,km\,s$^{-1}$ are not unexpected for shocks around expanding radio sources \citep{Bicknell1998,Axon1998,Capetti1999} and so it is plausible that adiabatic expansion could explain the decrease in flux of C3.

For the lower surface brightness components the characteristic radiation decay time-scale is 
$\sim$10$^{5}$ years, which is similar to the value estimated by \citet{Pedlar93}. However given our uncertainty in measuring the flux density changes we do not speculate further about these components.
Clearly, for C4, injection of new high energy particles dominates over any radiation or expansion losses.

\section{The possible connection between the radio jet  and the emission line region}
\label{sec:EmissionLines}

\begin{figure}
\begin{centering}
\subfloat[]{\includegraphics[width=\columnwidth]{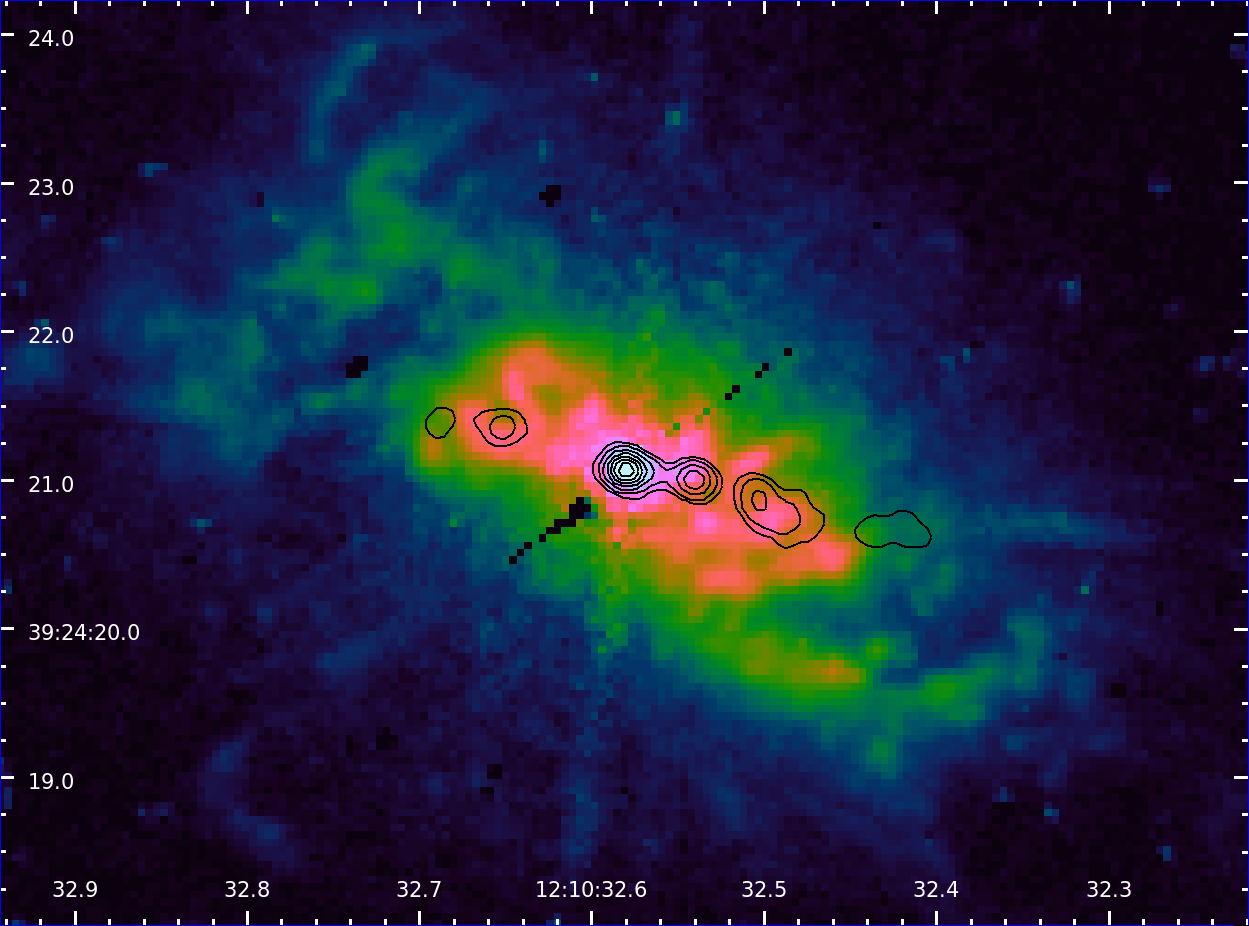}}\\
\hspace*{\fill}
\subfloat[]{\includegraphics[width=\columnwidth]{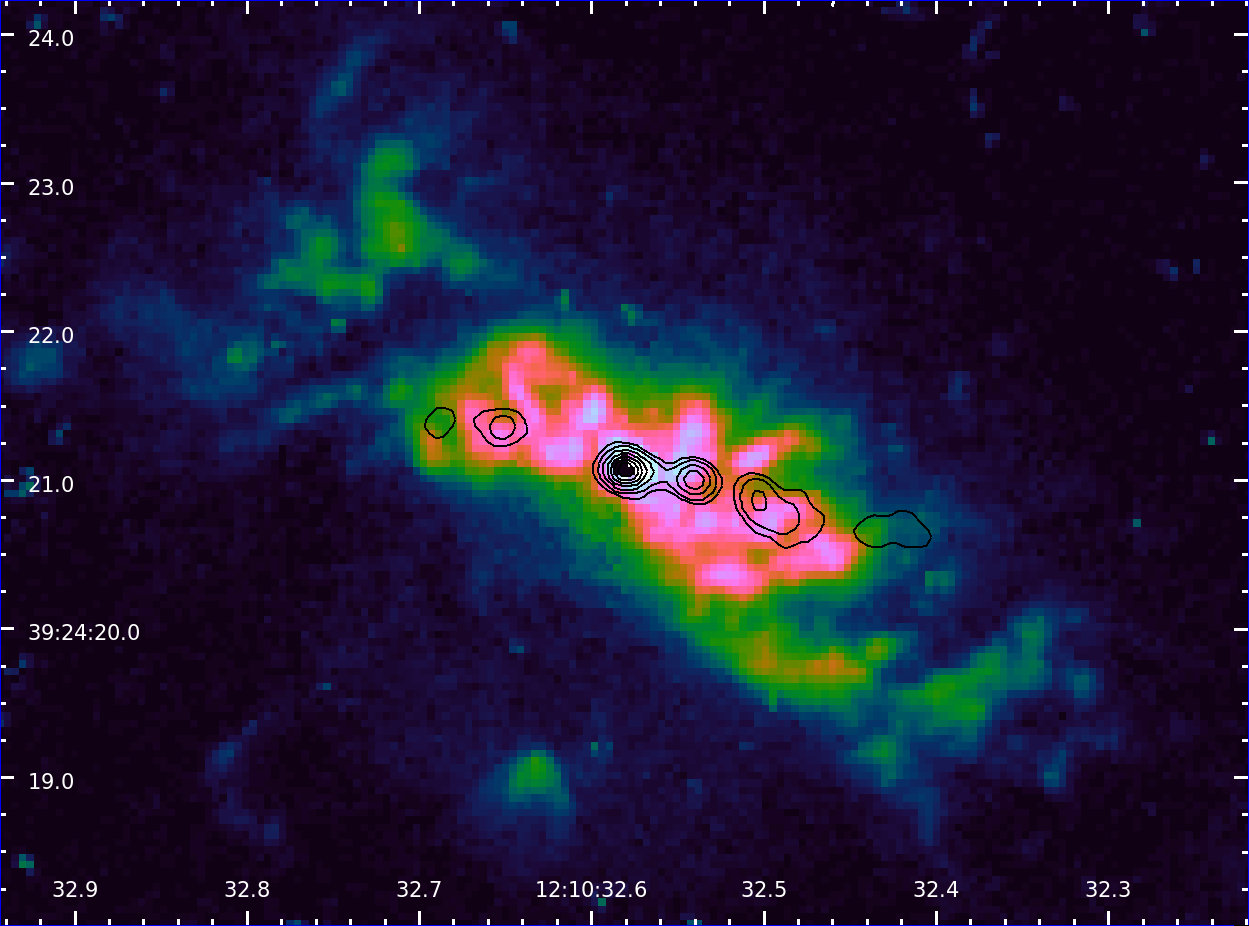}}\\
\hspace*{\fill} % separation between the subfigures
\subfloat[]{\includegraphics[width=\columnwidth]{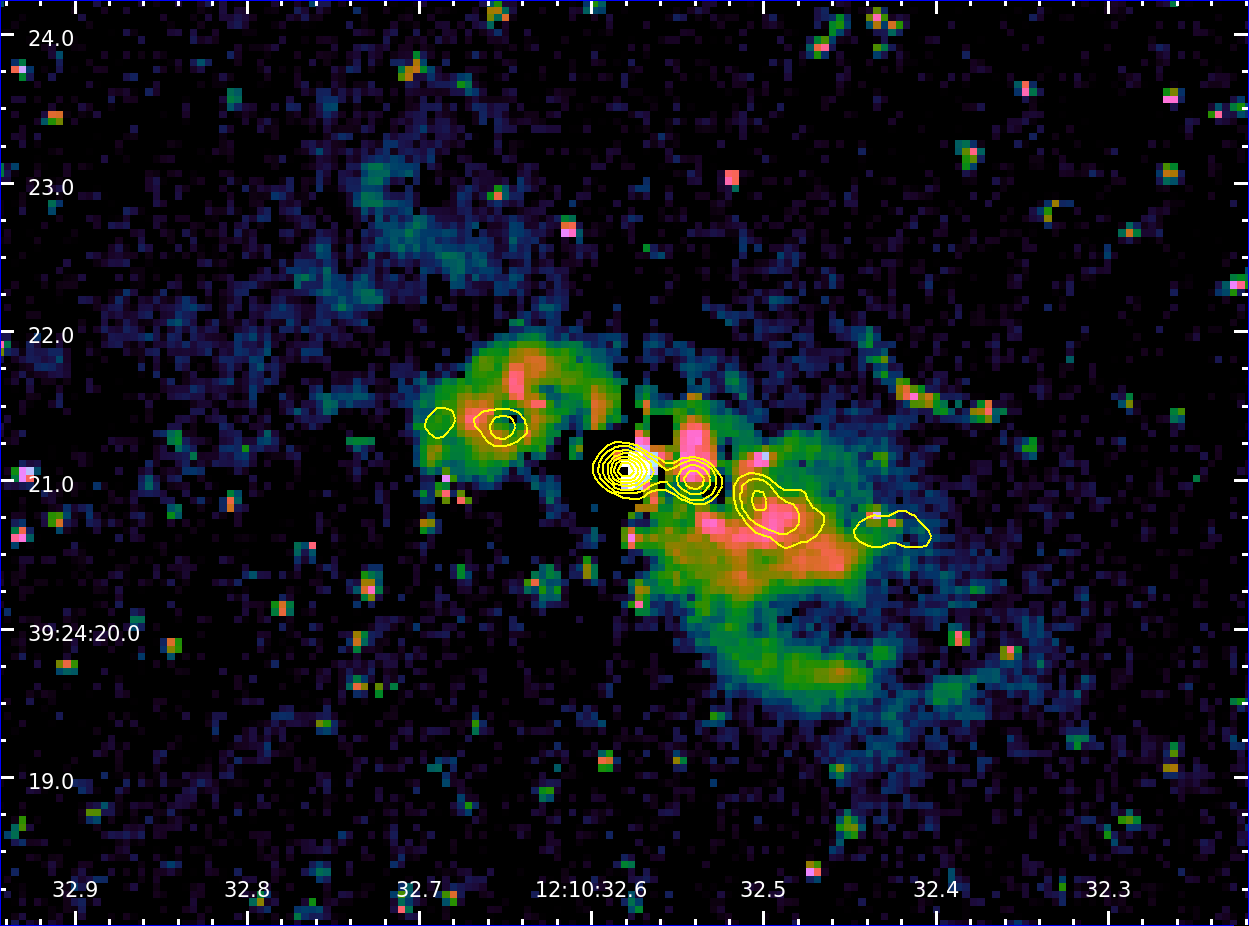}}\\
\hspace*{\fill} % separation between the subfigures
\caption{HST emission line images of NGC\,4151 for (a) H$\alpha$, (b) [O~III] and (c) [O~II]. The full resolution eMERLIN radio contours are plotted on top and are at 2, 4, 9, 16, 25, 36, 49, 64 mJy/beam. In all images, north is up and east is to the left. All three images correspond to the central $\sim$8 $\times$ 6$\arcsec$ ($\sim$740 $\times$ 550\,pc) of the nucleus of NGC\,4151.}
\label{fig:O2optical}
\end{centering}
\end{figure}

\begin{figure*}
	\includegraphics[width=0.95\textwidth]{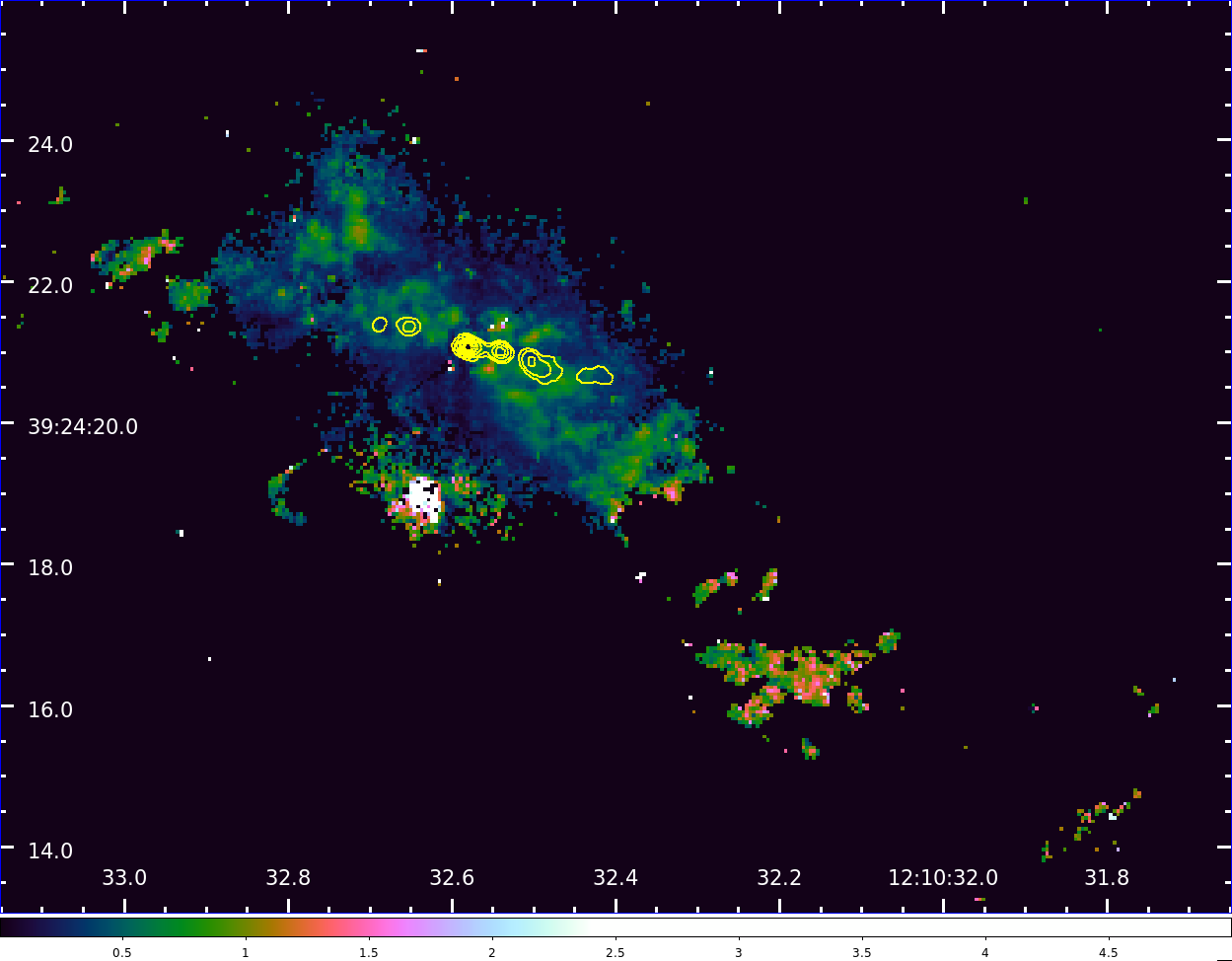}
    \caption{Image showing the ratio of the optical [O~III] to H$\alpha$ line emission from HST imaging of the central $\sim$21 $\times$ 12$\arcsec$ ($\sim$1.9 $\times$ 1.1\,kpc) of NGC\,4151. The full resolution eMERLIN radio contours are plotted on top and are at 2, 4, 9, 16, 25, 36, 49, 64 mJy/beam
We clipped out any pixels with S/N $<4$ to remove spurious artefacts. 
The larger-scale structure of the ionisation region is seen south west of the core. This lines up more closely with the PA of the overall line emission $\sim$50$^{\circ}$ but it is clear that the inner region bounded by the radio jet is at a slightly different angle at $\sim$57$^{\circ}$. In this image, north is up and east is to the left.}
    \label{fig:HaOIIsub}
\end{figure*}

In Fig.~\ref{fig:O2optical} we show the HST emission lines images from our new reduction in
H$\alpha$, [O~III], and [O~II] and in Fig.~\ref{fig:HaOIIsub} we show an emission line ratio image, [O~III]/H$\alpha$. eMERLIN radio contours are superimposed in all cases.

In the emission line images (Fig.~\ref{fig:O2optical}) there is a relatively bright region of emission of size $\sim8 \times 3\arcsec$ (730\,pc $\times$ 270\,pc) broadly surrounding the radio jet, although the 
emission line region (major PA $\sim$57$^{\circ}$) and jet (PA $\sim 79^{\circ}$) are not exactly aligned. The [O~III]/H$\alpha$ ratio image defines a larger cone-like structure (PA $\sim 53^{\circ}$) extending at least 11$\arcsec$ (1\,kpc) from the core. The emission line images show the presence of sub-structures, arches and haloes, indicating a clumpy distribution of the gas.

The range of [O~III]/H$\alpha$ ratio we show is between 0 and 5, based on values obtained for similar ELR for radio galaxies (Baldi et al., in prep). This ratio is a useful diagnostic tool for investigating the nature of the ionising source, i.e. shocks or photoionisation
\citep{Kewley06,Allen08,CapettiBaldi2011}. [O~III]/H$\alpha$<1 can be reproduced by shocks or by a weak radiation field from an AGN or star-formation, while strong illumination from the AGN or a young stellar population can yield [O~III]/H$\alpha$>2 (Baldi et al., in prep). 

The [O~III]/H$\alpha$ values increase from very low values ($\sim 0.5$) near the core and around the radio jet to $>1.5$ in the more distant parts of the cone. This change in ratio suggests a change in the mechanism of ionisation from shock heating near the core to AGN photoionisation further out. We therefore next consider in more detail whether shocks, such as might be associated with the jet, can explain the ionisation of the inner, brighter, ionisation region. The bright region 2$\arcsec$ (180\,pc) south of the core which is visible in the [O~III] and [O~III]/Halpha image is a ``ghost`` caused by internal reflection \citep{Hutchings98}.

\subsection{Origin of the emission line region: jet or AGN?}

The ratio of the radio luminosity to the emission line luminosities provide powerful diagnostics of the source of the ionisation, \citep[e.g.,][]{Pakull10}. \cite{Bicknell1998} present the [O~III] and radio luminosities of various samples of AGN. They state that the [O~III] luminosities are consistent with the predictions of a model for ionisation based on the expansion of radio lobes \citep{Bicknell1997}. Here we therefore calculate the emission line luminosities of the ELR.

To determine the H$\alpha$, [O~III] and [O~II] luminosities of the brighter extended ELR surrounding the radio jet we select a circle of radius 2$\arcsec$ (180\,pc), centred on C4. We performed a standard analysis, using the task {\scriptsize RADPROF} in IRAF \citep{IRAF1, IRAF2} for the aperture photometry of the \textit{HST} images F658N, F502N, F375N. To estimate the emission line fluxes from the ELR we must first measure and remove the large unresolved component from the AGN nucleus. We therefore measure the flux within a radius of 0.125$\arcsec$ (11\,pc) centred on the core. We subtract this flux from the flux within the 2$\arcsec$ radius area. We convert the extracted fluxes into physical units using the parameter {\scriptsize PHOTFLAM}, the flux-density normalisation value, and the image bandwidths. We correct the fluxes for reddening using the Calzetti dust extinction law \citep{CalzettiLaw}. Furthermore, we estimate the contamination from other emission lines falling within these bands. For F658N, to estimate the H$\alpha$ we take into account the [N~II] doublet, using the observed ratio of [N~II]$\lambda$6583 = 0.68 H$\alpha$ from the optical spectrum \citep{Ho97a}. Only the [O~III]$\lambda$5007 line of the [O~III] doublet falls in the F502N band and only the [O~II]$\lambda\lambda$,3726,3729 doublet falls in the F375N band. Hence, the total ELR fluxes within the 2$\arcsec$ radius but not including the unresolved contribution from the nucleus, are 1.54 $\times$ 10$^{-14}$, 1.10 $\times$ 10$^{-13}$ and 1.43 $\times$ 10$^{-13}$\,erg\,s$^{-1}$\,cm$^{-2}$, respectively for H$\alpha$, [O~III], and [O~II]. Assuming a distance of 19\,Mpc for our target, the calculated luminosities are 6.65 $\times$ 10$^{38}$, 4.74 $\times$ 10$^{39}$ and 6.16 $\times$ 10$^{39}$\,erg\,s$^{-1}$.  The ratio of the [O~III] to radio luminosity here is consistent with the observations presented by \cite{Bicknell1998}.
Changes in the positioning and size of the subtraction region of the central core is the major source of error in measurement of line fluxes from the ELR and may be up to a factor of 3.

Shock models can be used to predict line luminosities. \citet{Nelson2000} used the {\scriptsize MAPPINGS II} code to estimate the expected H$\beta$ luminosity from shocks in NGC\,4151. Here we repeat that analysis using results from the {\scriptsize MAPPINGS III} code \citep{DopitaSutherland,Allen08} similar to the analysis of eMERLIN and \textit{HST} observations of M~51b by Rampadarath et al. (submitted). First, taking H$\beta$/H$\alpha$=0.29 from long-slit spectra \citep{Ho97a}, then from our measured value of H$\alpha$ luminosity (above) we find that L$_{H\beta}$ = 1.93 $\times$ 10$^{38}$\,erg\,s$^{-1}$. The ratio of H$\beta$ luminosity to luminosity in ionising radiation produced from shocks is $\propto$ v$_{s}^{-0.59}$, where v$_{s}$ is the velocity of the shock. Here, in agreement with \cite{Nelson2000}, we find that a low velocity shock overpredicts L$_{H\beta}$ and we require v$_{s}$ $>$ 2700\,km\,s$^{-1}$, similar to the calculated jet expansion speed, to reduce the predicted L$_{H\beta}$ to the observed level. We note that the {\scriptsize MAPPINGS III} code assumes a relatively high conversion efficiency of jet kinetic power into ionising radiation ($\sim$ 27/77, \citealt{Weaver1977}). A reduced efficiency could therefore make our observed L$_{H\beta}$ consistent with lower shock velocities. We also note that \cite{Weaver1977} refers to low velocity stellar winds but faster shocks with velocities $>$ 1500\,km\,s$^{-1}$ are observed in narrow line regions of other nearby radio-bright Seyferts, similar to NGC\,4151 \citep{Axon1998,Capetti1999}. 

The emission line ratios [O~III]/H$\beta$ and [O~II]/H$\beta$ can provide another constraint on the shock velocities \citep{Allen08}. Our ratio values from the 2$\arcsec$ apertures are $\sim$25 for [O~III]/H$\beta$ and $\sim$30 for [O~II]/H$\beta$. These ratios imply v$_{s}$ $\gg$ 1000\,km\,s$^{-1}$, broadly consistent with the velocities derived from L$_{H\beta}$. 

We qualitatively compare the energetics from the AGN radiation with those of the radio jets. We estimate the kinetic jet power from the radio core luminosity according to the empirical relation found by \cite{Merloni2007} by using 5\,GHz core components from VLA data. For our target, we use the 5\,GHz data obtained in \citet{Pedlar93} and hence calculate a jet power of $1.4 \times 10^{42}$\,erg\,s$^{-1}$. We can estimate the AGN radiative power from the X-ray nuclear emission. \cite{Wang2011b} measure the nuclear X-ray emission using Chandra as 1.13 $\times$ 10$^{-10}$\,erg\,s$^{-1}$\,cm$^{-2}$, corresponding to a luminosity of $4.6 \times 10^{42}$\,erg\,s$^{-1}$, although they do note that this is variable. Therefore the values derived for the energetics of the AGN radiation and of the radio jets are consistent assuming uncertainties in the \cite{Merloni2007} relation.

\citet{Mundell03} concluded that the AGN radiation field is the main source of ionising power in the ELR, but that the radio jet interaction with line-emitting clouds could contribute to the observed ionisiation. By fitting Chandra X-ray spectra, \citet{Wang11} also argued that shocks could make up to 12 per cent of the ionised extended emission. Generally, our results are consistent with the conclusions of \citet{Mundell03} and \citet{Wang11} about the jet contribution to the ionisation of the ELR. Combined with changing ionisation parameter in our [O~III]/H$\alpha$ image, we therefore conclude that in the ELR at greater distances from the AGN core than the radio jet e.g. further away from C4 than 360\,pc, that photoionisation from the AGN is the only significant source of ionising power for the ELR. However in regions close to the radio structure ($\le$360\,pc), the jet cannot be ruled out as at least one component of the ionising power. As we cannot fully disentangle the contribution of the two main ionising sources, we cannot quantitatively estimate the fractional ionisation due to the radio jet relative to the AGN illumination in the ELR from our results.

\section{Summary and Conclusions}
\label{sec:SummaryandConclusions}

The Seyfert galaxy NGC\,4151, with twin radio jets, is the radio-brightest of the so-called 'radio-quiet' AGN and is thus one of the few such AGN for which it is possible to observe temporal changes in the structure of its jets. Here we present high-resolution 1.5\,GHz images obtained as part of the LeMMINGs legacy survey with the eMERLIN array. We compare these images with those made 22 years previously with the structurally very similar MERLIN array (M95). These images clearly show that the central AGN (component C4) has brightened by almost a factor 2. The components in the eastern jet (C5, C6) do not appear to have changed but components in the western jet, particularly C3, seem to have decreased in intensity. These observations show that the AGN core, C4, is still very much active in NGC\,4151 and still injecting particles into the inner jet but energy loss mechanisms are dominating further out in the jet.

We detect no significant change in the separation between the core, C4, and the only other unresolved component in the eMERLIN image, C3. The resultant upper limit to the jet velocity $\le$0.04c is consistent with the upper limits set by VLBI observations \citep{Ulvestad2005} over a shorter (4 year) time period. However, given the beam size of 150~mas, we do not claim any detectable movement of C3 away from the core C4.

We derived a spectral index image from within the eMERLIN bandwidth. The radio spectrum of the core, C4, is flat as expected from a jet base, and steepens down the jet, probably due to radiative losses. Further eMERLIN observations will be made at 5\,GHz in the near future which will better constrain the spectral indices of the jet components and resolve the smaller-scale emission.

We have estimated the magnetic fields in all components in NGC\,4151, assuming the minimum energy condition. The values found, taking sizes from eMERLIN observations, are $\sim$few mG, consistent with similar MERLIN observations in the
literature. Magnetic fields derived using VLBI observations are factors of 10 higher. The characteristic synchrotron decay time-scale, inferred from the magnetic field derived from VLBI observations for the unresolved bright component C3, is of the order of $\sim$700 years, longer than that required to account for the flux decrease in 22 years. However, we find that given adiabatic losses, a very small linear expansion factor of 6 per cent can produce the measured flux density decrease of 25 per cent.

We present newly-reduced high-resolution optical emission line images (H$\alpha$, [O~III], and [O~II]) from HST of the nucleus of NGC\,4151. The misalignment between the radio and optical line emission regions has been discussed by previous authors, e.g. \citet{Evans1993} and \citet{Pedlar93} and are not discussed further here. However we do use these data to investigate the origin of the ionisation of the ELR as a function of distance from the AGN. We note that the [O~III]/H$\alpha$ ratio close to the AGN, near the radio jet, is best produced by shocks whereas, far from the AGN ($\ge$360\,pc), the ratio is more consistent with photoionisation. We also note that the ratio of radio to emission line luminosity shown here is similar to that found by \citet{Bicknell1998} for various samples of radio galaxies and is consistent with models for ELR ionisation based on expanding radio lobes. Although our estimate is very uncertain, the derived expansion velocity of C3 is in agreement with the higher shock velocities needed to avoid overproduction of H$\beta$ emission.
We also note that the brightest parts of the ELR are those directly surrounding the radio jet. We therefore conclude that although the parts of the ELR further from the core than the eMERLIN radio jet ($\ge$360\,pc) are ionised largely by photons from the AGN, in the region of the ELR close to the radio jet, the jet is probably also a significant contributor to the ELR ionisation.

Although it is clear that there is a two-sided radio jet interacting with the ISM in NGC\,4151, it remains to be seen whether it is characteristic of all low-luminosity AGN or is in fact a curious, but intriguing exception to the rule \citep{Ulrich2000}. To explore the ISM-jet connection on a broader scale, we require systematic studies of large samples of low-luminosity AGN at sub-arcsec-resolution. Radio observations of nearby sources such as those currently being made with the LeMMINGs survey are able to resolve sub-kpc scale radio emission to investigate jet formation. However, only in conjunction with multi-wavelength data of comparable angular resolution (such as that obtained from \textit{HST} and \textit{Chandra}) can we learn about the physics of AGN feedback into the ISM.

\section*{Acknowledgements}

We thank the anonymous reviewer for their comments and revisions. We acknowledge funding from the Mayflower Scholarship from the University of Southampton afforded to DW to complete this work. This publication has also received funding from the European Union's Horizon 2020 research and innovation programme under grant agreement No 730562 [RadioNet]. IMcH thanks the Royal Society for the award of a Royal Society Leverhulme Trust Senior Research Fellowship. RDB and IMcH also acknowledge the support of STFC under grant [ST/M001326/1]. JHK acknowledges financial support from the European Union's Horizon 2020 research and innovation programme under Marie Sk\l{}odowska-Curie grant agreement No 721463 to the SUNDIAL ITN network, and from the Spanish Ministry of Economy and Competitiveness (MINECO) under grant number AYA2016-76219-P. DMF wishes to acknowledge funding from an STFC Q10 consolidated grant
[ST/M001334/1]. EB and JW acknowledge support from the UK's Science and Technology Facilities Council [grant number ST/M503514/1] and [grant number ST/M001008/1], respectively. FP has received funding from the European Union's Horizon 2020 Programme under the AHEAD project (grant
agreement No 654215). We also acknowledge Jodrell Bank Centre for Astrophysics, which is funded by the STFC. eMERLIN and formerly, MERLIN, is a National Facility operated by the University of Manchester at Jodrell Bank Observatory on behalf of STFC. Some of the observations in this paper were made with the NASA/ESA Hubble Space Telescope, and obtained from the Hubble Legacy Archive, which is a collaboration between the Space Telescope Science Institute (STScI/NASA), the European Space Agency (ST-ECF/ESAC/ESA) and the Canadian Astronomy Data Centre (CADC/NRC/CSA). DW would also like to thank Alessandro Capetti, Sam Connolly, Sadie Jones and Anthony Rushton for useful discussions.

%%%%%%%%%%%%%%%%%%%%%%%%%%%%%%%%%%%%%%%%%%%%%%%%%%



\bibliographystyle{mnras}

\bibliography{mybibliography.bib}

%%%%%%%%%%%%%%%%%%%%%%%%%%%%%%%%%%%%%%%%%%%%%%%%%%

%%%%%%%%%%%%%%%%% APPENDICES %%%%%%%%%%%%%%%%%%%%%

%\appendix

%%%%%%%%%%%%%%%%%%%%%%%%%%%%%%%%%%%%%%%%%%%%%%%%%%

% Don't change these lines
\bsp	% typesetting comment
\label{lastpage}
\end{document}